\documentclass[
 reprint,
 preprintnumbers,
 amsmath,
 amssymb,
 aps,
 prstab,
 floatfix,
]{revtex4-1}

\usepackage{graphicx}			
\usepackage{verbatim}			
\usepackage{wasysym}			
\usepackage{mathtools}			
\usepackage[makeroom]{cancel}		
\usepackage{dcolumn}			
\usepackage{hyperref}			
\usepackage{color}			

\newcommand{\pd}{\partial}				
\newcommand{\dd}{\mathrm{d}}				

\newcommand{\Nb}{N_{\text{p}}}		
\newcommand{\Qb}{ Q_{\beta}}		
\newcommand{\Qs}{ Q_s}		

\begin{document}

\title{Core-Halo Collective Instabilities}
\author{Alexey Burov}
\email{burov@fnal.gov}
\affiliation{Fermilab, PO Box 500, Batavia, IL 60510-5011}
\date{\today}

\begin{abstract}
At strong space charge, transverse modes of the bunch core may effectively couple with those of the halo, leading to instabilities well below the core-only transverse mode-coupling threshold. 
\end{abstract}


\maketitle

 

{\it Introduction.}

Collective instabilities limit intensity of charged particle beams in many accelerators. Such instabilities, caused by mutual interaction of the beam particles, lead either to density degradation or to beam loss, full or partial. Both, Coulomb fields of the bunch space charge (SC) and wake fields left behind the particles, are generally important in this respect. While the former change the spectra of the individual and collective modes, making them more or less prone to instabilities, the latter, being non-Hamiltonian, are able to drive the instabilities. 

Hereafter, a beam in a circular machine is modeled as a single bunch, with short-range wakes only (no multi-turn wakes), and with fully linear focusing. When wake and SC fields can be neglected, frequencies of the bunch transverse collective modes represent a series of equidistant sidebands around the main transverse frequency, the betatron frequency $\omega_\beta$. Distances between the neighbor sidebands are equal to the longitudinal frequency of the bunch particles inside the potential well, the synchrotron frequency $\omega_s$. In reference to the betatron frequency, frequencies of the transverse collective modes $\Omega_k$ represent a series, $\Omega_k=k \omega_s$, where integers $k=0,\; \pm1,\; \pm2,\,...$ serve as the mode numbers. For circular accelerators, it is conventional to normalize all the beam frequencies to the revolution frequency; the resulting values are called {\it tunes}. However, for analysis of the collective motion, the revolution frequency does not mean much; it makes more sense to normalize all the sidebands $\Omega_k$ to the synchrotron frequency instead, dealing with properly normalized collective tunes $\nu_k=\Omega_k/\omega_s$. With negligible interaction between the particles, the collective tunes $\nu_k=k$; however, SC and wake both shift the tunes from these unperturbed values.    

The wake force $F$ is conventionally represented by the wake function $W(s)$ between two unit charges separated by a distance $s$: $\; F(s) = W(s)\,y/C_0$, where $y$ is an offset of the leading particle and $C_0$ is the ring circumference, see e.g.~\cite{chao1993physics}. With these forces and without SC, the bunch particles represent a set of identical harmonic oscillators, acting on each other by linear forces $F(s)$. The wakes are normally causal, i.e. the trailing (or tail) particle does not act back on the leading (or head) one, $F(s)=0$ at $s<0$, so Newton's third law does not apply; wakes make the system non-Hamiltonian. Nevertheless, this sort of bunch is stable below a certain threshold intensity, since the equations of its collective dynamics reduce to an eigensystem problem of a real non-degenerate matrix; at zero wake it is diagonal and filled with consecutive integers $k=0,\; \pm1,\; \pm2,\,...$. The collective tunes are given by the eigenvalues of that matrix. With a non-zero wake, some of the eigenvalues may become complex, coming in pairs: if $\nu_k$ is the eigenvalue, then its complex conjugate $\nu_k^*$ is the eigenvalue too. Due to this fact, the complex tunes with $\Im \nu \neq 0$ may appear only after coupling of two modes, whose real tunes become identical at the threshold and get opposite imaginary parts only above it. That is why the instability is called TMCI, the transverse mode coupling instability.

With all the collective frequencies measured in units of the synchrotron frequency, the same rule applied to the SC frequency shift $\Delta \omega_\mathrm{sc}$ requires the introduction of the space charge parameter $q = \Delta \omega_\mathrm{sc}/\omega_s$ as the measure of its strength. The wake parameter $w$ can be introduced in a similar way, as $\sim N_p\,W_0 \max_k d\nu_k/d(N_pW_0)$, where $W_0$ is an amplitude of the wake function, $N_p$ is the number of particles per bunch, and the derivative is taken at zero intensity; in other words, the wake parameter represents the maximal collective tune shift in the units of the synchrotron frequency. 

Influence of SC on the TMCI was considered in a series of publications, starting from a pioneer work of M.~Blaskiewicz~\cite{blaskiewicz1998fast}, followed by more detailed analysis of Refs.~\cite{burov2009head, PhysRevAccelBeams.20.114401, balbekov2017transverse, Zolkin:2017sdv}. It has been shown that the wake threshold $w_\mathrm{th}$ almost always grows linearly with SC, 
\begin{equation}
\label{TMCIth0}
	w_\mathrm{th} \propto q \,,
\end{equation}
as soon as SC is strong, $q \gg 1$.  For years, nobody published anything about a certain strangeness of this stability condition: its total insensitivity to the number of particles. Since both terms of Eq.~(\ref{TMCIth0}) are proportional to the bunch intensity $N_p$, the latter simply cancels out. Thus, were it correct, such a beam with sufficiently strong SC, being stable at some intensity, would remain stable at higher intensity, regardless of how much! This fantastic conclusion followed from strict and thorough independent analyses by the authors mentioned above in this respect. 

Resolution of the conundrum was recently suggested in Ref.~\cite{Burov:2018pjl}. 
According to that, a mistake, or rather unreflected prejudice, of the author himself and the related community consisted in a tacit equating of instabilities in general with {\it absolute instabilities}, i.e. instabilities in general were equated with the existence of a collective mode with the positive imaginary part of the tune, $\Im \nu > 0$. Well, this seemingly obvious equality is deceptive. Collective instabilities do not reduce to absolute ones; this family of beasts includes another genus as well, namely, {\it convective instabilities}. 

According to Landau and Lifshitz~\cite{LandLifHydro}, a distinction between the two genera of instabilities can be described as follows. For an absolute instability, initial perturbations cause an unrestricted exponential growth everywhere in the medium; to suppress it, a damping rate must exceed the nonzero growth rate. For the convective instability, instead, there is only a spacial amplification, and the perturbation eventually decays everywhere when a dissipation is added, no matter how tiny. Generally speaking, convective instabilities are not less dangerous than absolute ones. Even if all the modes are stable in the absolute sense, convective amplifications along the bunch can be so large, that for all practical purposes the beam would and should be considered unstable. Due to the absolute stability, these convectively unstable modes have their amplitudes still bounded; that is why such modes may be characterized as {\it saturating convective instabilities}, or SCI, in distinction with {\it unbounded convective instabilities}, UCI; the latter are known in linacs as beam breakups.  Resolution of the conundrum consists in a demonstration that at strong SC, there is a significant interval of the wake parameters corresponding to an absolutely stable and convectively unstable bunch, with large head-to-tail amplifications. The lower limit of this interval is about the same as the TMCI threshold at zero SC, $w^0_\mathrm{th}$, while the upper limit is the TMCI threshold $w_\mathrm{th}(q)$ at the given SC parameter $q \gg 1$; typically $w_\mathrm{th}(q)/w^0_\mathrm{th} \simeq q$. Since the amplification coefficient of the SCI depends exponentially on the wake parameter, the amplification quickly reaches a practically intolerable level of hundreds and thousands. A convective instability with large amplification can be considered as a special metastable state: even a tiny, otherwise totally negligible, tail-to-head feedback, provided by a multi-revolution or a coupled-bunch wake could be sufficient to drive an absolute instability. Thus, the latter can be called absolute-convective instability, ACI. With sufficiently large amplification, any damper, including the conventional resistive one, turns, in principle, into an ACI generator. 

Keeping all this in mind, we still may imagine a bunch without all these feedbacks, injection errors and aperture limitations, and ask the question: could there be anything at all in the bunch itself that can still limit the amplifications of the convective instabilities at wake parameters considerably below the TMCI threshold at strong SC?

{\it Core-Halo Interaction.} Before trying to answer that question, let us show what sort of reason is behind the strange stability condition~(\ref{TMCIth0}); what features of SC provide such a dramatic elevation of the TMCI wake threshold, as it suggests. Well, this SC ability is caused by a fact that both SC and wake tune shifts are typically of the same sign; they are both defocusing. At strong SC, all the modes with intra-slice motion are strongly detuned from the rigid-slice modes, mostly coupled with wake; modes of the opposite groups cannot cross, and modes with non-rigid slices are insensitive to wake. Another important point is that typically wake mostly shifts (down) the mode $0$, responsible for bunch motion as a whole, while other modes are shifted to a lesser extent, thus wake works on the mode divergence. In short, that is why TMCI {\it vanishes} at strong SC: the negative modes with considerable intra-slice motion are SC-separated from $0^\mathrm{th}$ and positive modes, while wake separates positive modes even more. If the modes cannot couple and there are no feedbacks, absolute instabilities are impossible. 

At this point, however, we may recall that even at strong SC the described reasoning breaks for some particles with large individual amplitudes, the {\it halo} particles, which feel SC to lesser extent. Although the relative number of such particles may be small, this small percentage could be compensated by a large amplification coefficient of the core convective instability. An important feature of the halo is that it is sensitive to wake but less sensitive to SC, and its reduced sensitivity to SC is of a different sort than that of the core. Specifically, the rarified halo practically does not feel its own SC, only that of the core, so all of its modes can be equally easily or non-easily excited by the motion of the core; essentially they are similar to no-SC modes, just shifted down a bit according to the halo's reduced SC tune shift, $q_h \ll q \equiv q_c$. Nothing essentially prevents coupling between the core and the halo modes, which effect can be dramatically enhanced by SCI of the core. Thus, we are coming to the idea of the {\it core-halo} mode-coupling instability at strong SC as an absolute instability, which wake threshold may be well below the conventional core-only TMCI threshold. To check this idea, we need a reasonable quantitative model of the bunch collective dynamics, where both core and halo modes can be taken into account.

{\it ABS Model.}
For that matter, our single bunch in a circular accelerator can be modeled similarly to Ref.~\cite{Burov:2018pjl}, assuming the same airbag square well (ABS) model of M.~Blaskiewicz~\cite{blaskiewicz1998fast}. The longitudinal potential well, which keeps the beam bunched, is assumed to be square, while all the bunch particles are supposed to have the same synchrotron (longitudinal) frequency. The only novelty we have to introduce here with this model is to represent the bunch as consisting of two fractions, core and halo, with different SC tune shifts and different populations. Before doing this, though, let us write down the conventional core-only ABS equations of motion in the form of Ref.~~\cite{Burov:2018pjl}: 
\begin{equation}
\begin{split}
& \frac{\pd\,x}{\pd \theta} + \frac{\pd\,x}{\pd \psi} =  i q (x - \bar{x}) + i\,F\, ;	\\
& F(\psi) = w \int_0^{|\psi|} \frac{ \dd \psi'}{\pi} W(\psi'/\pi) \bar{x}(\psi-\psi') \, ;\\
& \bar{x}(\psi) \equiv x(\psi)/2 + x(-\psi)/2 \,.\\
\end{split}
\label{MainX}
\end{equation}
Here $x=x(\theta,\psi)$ is a slow amplitude of the transverse oscillations at time $\theta$, measured in the synchrotron radians, and of the synchrotron phase $\psi$; the searched-for function $x$ is periodical on the latter variable, supposed to change from $-\pi$ to $0$ for the tail-to-head moving particles, or for the $+$ flux, and from $0$ to $\pi$ for the $-$ flux, moving back to the tail; the bunch length is 1 in these units. The term $\propto q$ is the SC force, and $F$ is the wake force with $W(s)$ as a dimensionless wake function; $\bar{x}(\psi)$ is the local centroid. The dimensionless wake parameter $w$ keeps in itself all dimensional values of the problem; it is defined as 
\begin{equation}
\label{WParam}
	w = \frac{\Nb W_0 r_0 R_0}{4\,\pi\,\gamma\,\beta^2\,\Qb \Qs}\,,
\end{equation}
with $\Nb$ as the number of particles, $W_0$ as the amplitude of the wake function in conventional units of Ref~\cite{chao1993physics}, $r_0$ as the particle classical radius, $R_0$ as the average radius of the machine, $\gamma$ and $\beta$ as the relativistic factors, $\Qb$ and $\Qs$ as the conventional betatron and synchrotron tunes. 

Due to its phase space periodicity, the amplitude $x$ can be Fourier-expanded over the phase $\psi$, leading to a set of ordinary homogeneous linear equations on the harmonics $A_n \equiv (2\pi)^{-1}\int_{-\pi}^{\pi} x \exp(-in\psi)\,\dd \psi$: 
\begin{equation}
\begin{split}
& i\dot{A}_n = n A_n - q(A_n - \bar{A}_n) - w \sum_{m=-\infty}^{\infty} U_{nm} \bar{A}_m \,; \\
& U_{nm} \equiv  {\int_0^1 \dd s \int_0^s \dd s' W(s-s') \cos(\pi ns) \cos(\pi m s')}\,,
 \end{split}
\label{Aeq}
\end{equation}
with the centroid's harmonic $\bar{A}_n=(A_n+A_{-n})/2$. This form of the dynamic equations makes its generalization for any number of bunch fractions fairly obvious: the Fourier amplitude of every fraction satisfies the same Eq.~(\ref{Aeq}) with its own relative shift of the synchrotron frequency $\Delta \omega_s/\omega_s$ and its own SC tune shift, i.e. with $n \rightarrow n (1+\Delta \omega_s/\omega_s)$ and its own parameter $q$. Each fraction, weighed with its relative intensity, contributes to the centroid $\bar{A}$. As a result, the problem can be solved with a reasonable accuracy in a reasonable CPU time, leading to a straightforward analysis of continuous and discrete van Kampen spectra~\cite{VanKampen:1955wh}, similar to Refs.~\cite{Burov:2010zz,Alexahin:2017fuq}. Here, however, a different approach is suggested and realized.

{\it Border of the Halo.}
Instead of presenting the bunch by a sufficiently large number of fractions, let us consider it as consisting of just two: the core and the halo, marked with the corresponding indices $c$ and $h$: 
\begin{equation}
\begin{split}
& i\dot{A}_n^c = n A_n^c +G^c_n - w \sum_{m=-\infty}^{\infty} U_{nm} \bar{A}_m \,; \\
& G^c_n \equiv -q_c(A_n^c - \bar{A}_n^c)-q_h p_h(A_n^c - \bar{A}_n^h)\,; \\
& i\dot{A}_n^h = n A_n^h + G^h_n - w \sum_{m=-\infty}^{\infty} U_{nm} \bar{A}_m \,;\\
& G^h_n \equiv - q_h(A_n^h - \bar{A}_n^c)\,; \\
& \bar{A}_n \equiv \bar{A}_n^c +  p_h \bar{A}_n^h \,,
\end{split}
\label{ACH}
\end{equation}
where $p_h \ll 1$ is a relative population of the halo. For zero wake, Eqs.~(\ref{ACH}) describe a Hamiltonian system, stable for any choice of the remaining parameters. 

The suggested two-fractional bunch model meets an obvious question: how to define the halo tune shift $q_h$? Where should the borderline between the core and the halo be drawn? Seemingly, there is no border in reality; thus, wouldn't whatever border of the model be as arbitrary and artificial as any other? This paper suggests a {\it natural} solution to this problem: let the system itself make the decision! For the given wake and (core) SC parameters, $w$ and $q \equiv q_c$, let the halo relative SC tune shift $\tilde{q} \equiv q_h/q_c$ be a free variable, and let us then find such a value for it that corresponds to the highest instability growth rate. That special value of the halo SC tune shift would point to the most effective, most powerful, and hence most important collective interaction, motivating to take such halo border as the {\it natural} border. With this dynamic definition, the border would be a function of the wake and SC parameters. The above implies that the halo relative population $p_h$ is a known function of its parameter $\tilde{q}$; indeed, as soon as the bunch distribution function is given, the population function can be computed. For a transversely Gaussian bunch inside the longitudinal square well, the partial SC tune shift $\tilde{q}$ versus two transverse actions $J_{x,y}$ was found by Lopez~\cite{Lopez:1993yta}:
\begin{equation}
\label{Qsc}
\tilde{q}_L=\int_0^1 \dd z \frac{\mathrm{I}_0(J_x z/2)-\mathrm{I}_1(J_x z/2)}{\exp(J_x z/2+J_y z/2)}\mathrm{I}_0(J_y z/2).
\end{equation}
Here, the $x$ direction is the one of the tune shift $q$; the actions $J_{x,y}$ are measured in units of their beam-average values, or emittances, assumed to be equal here; $\mathrm{I}_{0,1}$ are the modified Bessel functions. Now, the portion of particles $p(\tilde{q})$ whose tune shifts do not exceed $\tilde{q}$ is obtained right away: 
\begin{equation}
\label{pq}
p(\tilde{q})=\int_0^{\infty}\int_0^{\infty} \dd J_x \dd J_y \mathrm{\Theta}(\tilde{q}_L(J_x,\,J_y)-\tilde{q}) e^{-J_x - J_y} \,,
\end{equation}
where $\mathrm{\Theta}$ stands for the Heaviside theta-function. For these general core-halo considerations, an approximation of large actions suffices, leading to $\tilde{q}_L \simeq 3/(2J_x+J_y)$, and, further, to the asymptotic expression of the halo population function,
\begin{equation}
\label{pqas}
p(\tilde{q}) \simeq 2 \exp(-1.5/ \tilde{q}) \,.
\end{equation}
At a sufficiently small $\tilde{q}$, the distribution $p$ is sharp; its relative width $\delta \tilde{q}/\tilde{q} \simeq 0.7\tilde{q}$; this fact supports the idea to consider all the halo particles as having the same SC tune shift, as soon as the halo border is sufficiently far. 
\begin{figure}[h]
\includegraphics[width=\linewidth]{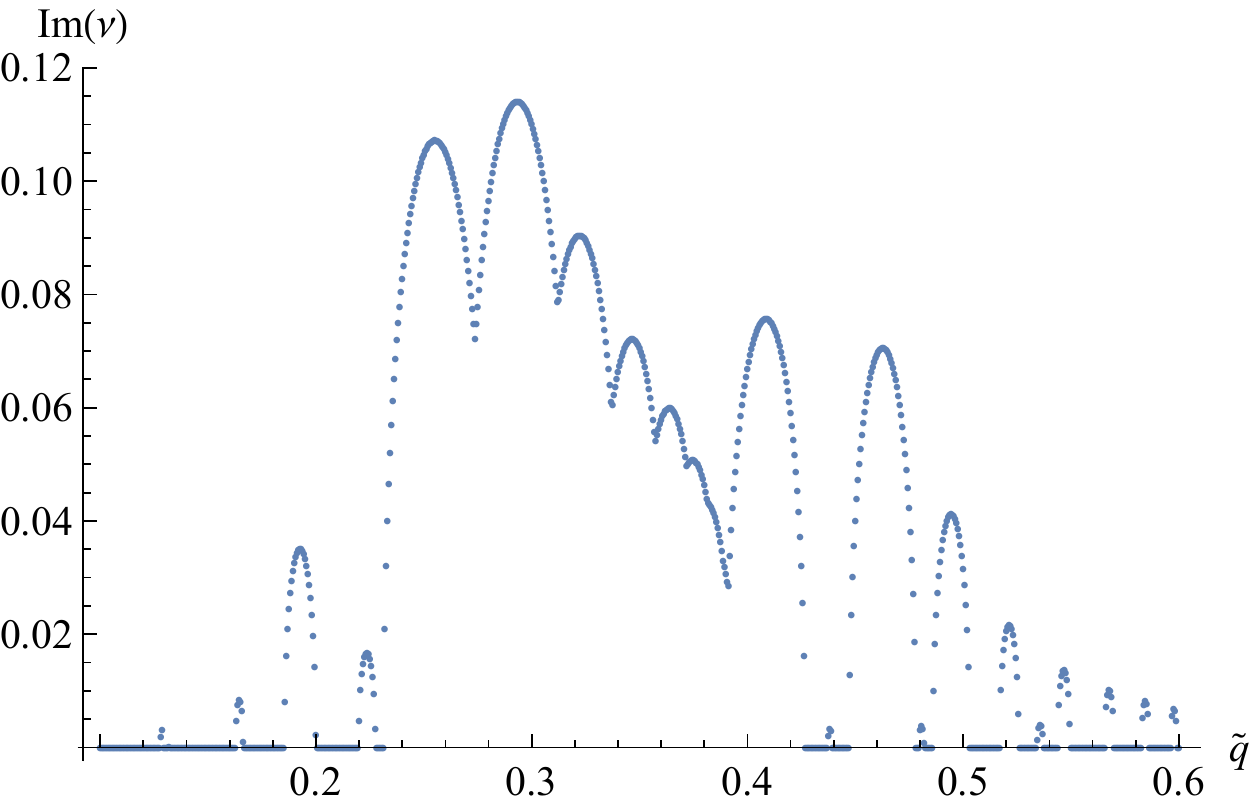}
\caption{\label{GRateQhrQW}
	Instability growth rate versus the halo parameter $\tilde{q}=q_h/q_c$ for the SC and wake $q_c=10$ and $w=4.$  
	}
\end{figure}
\begin{figure}[h]
\includegraphics[width=\linewidth]{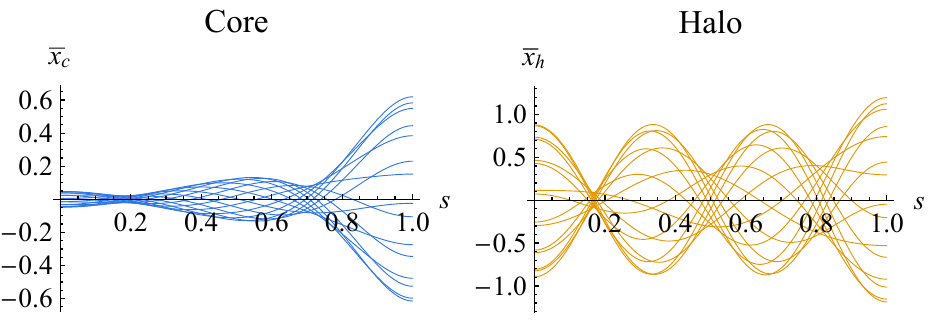}
\caption{\label{StroboQ10W4Qhr0p29}
	Centroid stroboscopic images of the core and halo components of the most unstable core-halo mode for the same $q$ and $w$ as in Fig.~\ref{GRateQhrQW}, at the most unstable $\tilde{q}=0.29\,$. Waists instead of nodes in the halo image tell about an absolute instability.
	}
\end{figure}
\begin{figure}[h!]
\includegraphics[width=\linewidth]{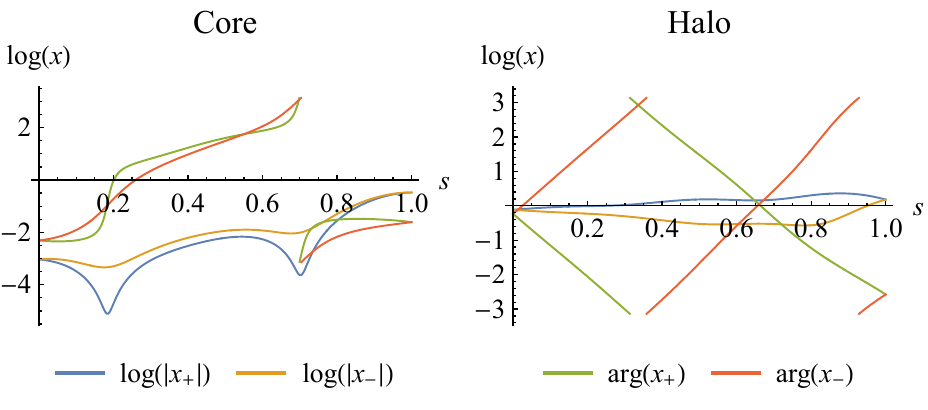}
\caption{\label{LogQ10W4Qhr0p29}
	Amplitudes and phases for the two fluxes of the core and the halo for the same modes, $+2$ and $+3$ correspondingly, as in Fig.~\ref{StroboQ10W4Qhr0p29}. The core mode is convectively unstable, with its $+$ and $-$ fluxes in phase, while the halo mode is similar to a typical no-SC modes having the $+$ and $-$ phases steadily running with opposite signs.
	}
\end{figure}

\begin{figure}[h]
\includegraphics[width=\linewidth]{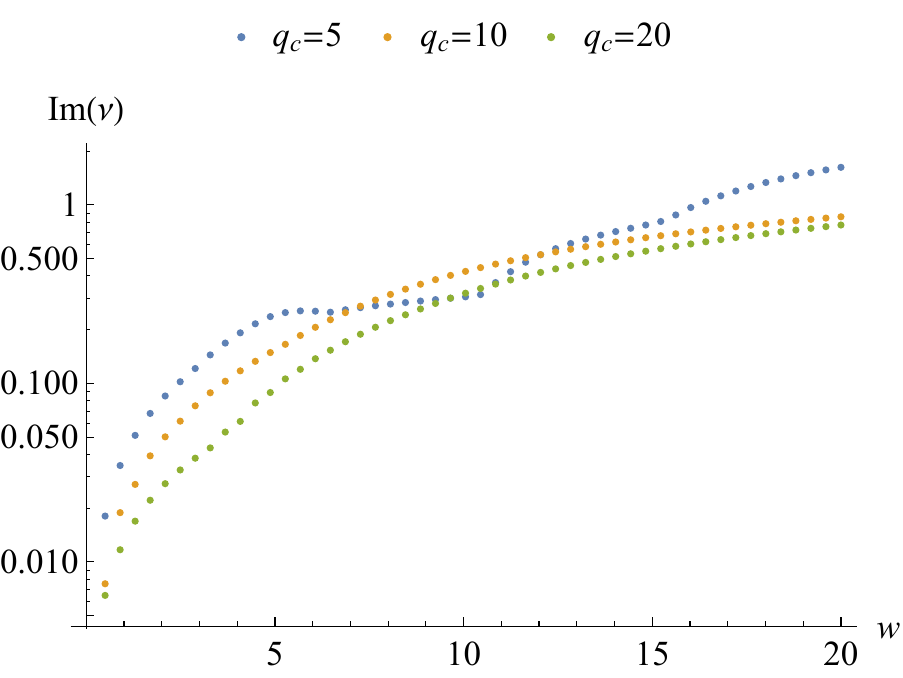}
\caption{\label{GratesWPlot}
	Growth rates of the most unstable modes versus wake parameter for three different SC parameters. Note the conventional TMCI threshold for $q_c=5$ at $w \approx 15$. 
	}
\end{figure}

{\it Results.}
To demonstrate the general properties of the instability, the simplest model wake function is chosen, the Heaviside theta, $W(s)=\Theta(s)$. After that, the eigensystem problems of Eqs.~(\ref{ACH}) are solved for the given wake and SC parameters, with the halo parameter $\tilde{q}$ initially free, but later assigned its {\it natural} value.

Figure \ref{GRateQhrQW} shows an example of how the calculated growth rate of the most unstable mode may depend on the halo parameter; the exemplifying wake is $\sim 4$ times above its no-SC TMCI threshold, and almost an order of magnitude below its TMCI threshold at this SC. The sequence of partially overlapped resonances of the core and halo modes is dominated by the resonance with the center at $\tilde{q} \approx 0.29$, the latter corresponding to the halo relative population $p_h \approx 1\%$. For the lower values of $\tilde{q}$, the halo would be too tiny, while for its higher values, SC depression of the mode coupling would be stronger; that is why the maximally unstable halo parameter $\tilde{q}$ is somewhere in between but not too close to $0$ and $1$. Figure~\ref{StroboQ10W4Qhr0p29} demonstrates stroboscopic images of the core and halo centroids of the most unstable eigenvector of Eqs.~(\ref{ACH}) at these parameters, $\Re (\bar{x}_c \exp(-i l \phi))$, $l=0,1,2,...$, taken with an arbitrary phase advance $\phi$, as if the oscillating centroids were observed a certain number of times, revolution after revolution, and the images superimposed. Figure~\ref{LogQ10W4Qhr0p29} suggests another representation of the same modes, showing the absolute values and complex arguments of the positive and negative fluxes for their core and halo components. The plots show that the core mode $+2$ is coupled with the halo mode $+3$. At this core and halo SC parameters, the tune of the halo mode $\nu^h_3 \approx 3-q_h=0.1$, which indeed is very close to the tune of the core mode $+2$, computed with the core-only model. The core mode is convectively amplified, showing the typical cobra shape and ACI phases in Fig.~\ref{LogQ10W4Qhr0p29}. This figure shows the halo mode with almost constant amplitude and steadily running phase, which is typical to no-SC modes; its waists instead of nodes at the right part of Fig.\ref{StroboQ10W4Qhr0p29} indicate an instability. 

Generally speaking, the halo's ability to play a feedback role is reduced by its low population; however, amplification of the core mode, considerable wake parameter and some halo SC tune shift enhance the core's sensitivity to the halo's oscillations, and thus may restore this ability to a level sufficient for driving the ACI.

Growth rate of the most unstable mode versus wake, with the natural halo parameters $\tilde{q}$, is presented in Fig.~\ref{GratesWPlot} for three values of SC. Growth rates that are too small at $w \lessapprox 1$ likely exceed the model accuracy, and should be rather considered as indistinguishable from zero; apparently, the instability threshold cannot be correctly predicted by this core-halo model, which accuracy is limited by the simplification of the two-fractional approach. What is clear though, is that the instability is already significant for the wake parameters well below the TMCI threshold. Note that the latter is clearly seen for $q=5$ at $w \approx 15$, in agreement with the no-halo calculations~\cite{blaskiewicz1998fast}.

The core-halo mode coupling suggests a new type of collective instabilities of bunched beams at strong SC, combining in themselves features of the TMCI and ACI. Contrary to the pure convective modes of the core itself, this instability is absolute; it may lead to the halo loss and the core emittance growth. On the other hand, the core-halo instability may be less limiting than the pure convective instabilities of the core, which typically have larger amplifications. 

{\it Acknowledgements.}
I am thankful to Yuri Alexahin for letting me know that in his computations with rather strong SC, as in Ref.~\cite{Alexahin:2017fuq}, the instability thresholds were found to be significantly lower when the rigidity of the bunch slices was not forced.   

Fermilab is operated by Fermi Research Alliance, LLC under Contract No. DE-AC02-07CH11359 with the United States Department of Energy.



\bibliography{bibfile}			

\end{document}